\documentclass[11pt]{article} 

\usepackage[utf8]{inputenc} 

\usepackage{amsmath}

\usepackage{caption}
\usepackage{subcaption}

\usepackage{geometry} 

\usepackage{graphicx} 

\usepackage{booktabs} 
\usepackage{array} 
\usepackage{paralist} 
\usepackage{verbatim} 

\usepackage{booktabs}
\usepackage{hyperref} 

\usepackage{placeins}

\usepackage{fancyhdr} 
\usepackage{sectsty}
\allsectionsfont{\sffamily\mdseries\upshape} 

\usepackage[nottoc,notlof,notlot]{tocbibind} 

\usepackage{hyperref}

\newcommand{\be}{\begin{equation}}
\newcommand{\ee}{\end{equation}}

\title{Bayesian Trading Cost Analysis and Ranking of Broker Algorithms}
\author{
    Vladimir Markov\\
    Bloomberg  L.P. in New York, NY\\
    \texttt{vmarkov2@bloomberg.net}
}
\date{}
\begin{document}
\maketitle
\begin{abstract}
We present a formulation of transaction cost analysis (TCA) in the Bayesian framework for the primary purpose of comparing broker algorithms using standardized benchmarks. Our formulation allows effective calculation of  the expected value of trading benchmarks with only a finite sample of data relevant to practical applications.
We discuss the nature of distribution of implementation shortfall,  volume-weighted average price, participation-weighted price and short-term reversion benchmarks. Our model  takes into account fat tails, skewness of the distributions and heteroscedasticity  of benchmarks. The proposed framework allows the use of  hierarchical models to transfer approximate knowledge from a large aggregated sample of observations to a smaller sample of a particular algorithm. 
\end{abstract}

\section{Introduction}

The performance of an investment is strongly related to execution costs of the investment. Often the transaction costs involved in trading securities may be large enough to substantially reduce or even eliminate the return on an investment strategy.   Transaction cost analysis (TCA) studies the relation between  execution costs (price impact model) and a set of stock and  order characteristics such as average daily trading volume, order size, market cap, volatility, spread, participation rate momentum, trading strategy and presence of limit prices. 

The precision of  the cost model depends on many factors and an important one is the sample size. If one is interested in ranking and comparing algorithms of different brokers, the data size quickly decreases because each broker has multiple execution algorithms.  Rigid statistical modeling is required in this situation. The robust market impact estimator should take into account peculiarities of execution benchmark distributions like the shape of the distribution, fat tails, skewness and  heteroscedasticity. The output of rigid statistical modeling is rewarding as it allows traders to compare brokers or a broker to evaluate the effect of a change in an algorithm. 


\section{Execution Algorithms Ecosystem}

Execution algorithms can be divided into two groups: schedule-based and liquidity-
seeking algorithms. The primary schedule-based algorithms are Time-Weighted Average Price (TWAP), Volume-Weighted Average Price (VWAP), Implementation Shortfall (IS) and Percentage of Volume (POV). They seek to meet or beat a well-defined
benchmark. In contrast, liquidity-seeking algorithms opportunistically search for mostly
dark and block liquidity, generally without targeting a specific benchmark. The choice of VWAP, POV or IS strategies is often dictated by a trader's
personal preferences and his view on liquidity extraction, investment constraints, portfolio manager instructions and an order's specifics related to the type and time horizon of
the investment strategy.

In an ideal world, the VWAP strategy minimizes market impact costs [Kissel and Glantz, 2003] with guaranteed completion of the order. Exposure to market risk  is controlled by the duration of the execution.

For traders and portfolio managers utilizing a mean-variance optimization approach, IS algorithms balance estimated market impact and market risk. 

The POV algorithm follows fixed volume participation. POV strategy doesn't guarantee order completion and has the least realized uncertainty of the schedule-based strategies as its prediction horizon is very short. POV guarantees to execute only what is possible to execute at reasonable cost. POV is used often because of its adaptivity, simplicity and the easy interpretation of its parameters.

Large institutional orders are often sent to opportunistic liquidity-seeking algorithms that harness liquidity from exchanges and dark pools [Markov and Ingargiola, 2013]. 

Given market condition and order requirement, selecting the right algorithm and tuning its parameters could significantly affect execution performance [Liu  and Phadnis, 2013]. The next-generation execution algorithms can be purely data driven [Bacoyannis et al., 2018].  

\section{Execution Benchmarks}
In the case of institutional trading,  an order (order level) is submitted by a portfolio manager to a trading desk.  A trader splits the order into one or several placements (placement level), which are then sent to designated brokers. 
On the placement level,  the performance of an algorithm can be quantified by four benchmarks: IS, VWAP, PWP and short-term reversion (see Appendix D). On the individual execution level: cost of non-execution and  effective spread capture metrics quantify the performance of limit order tactics, while the  price impact of individual trades quantifies  the performance of market order tactics.  Venue analysis is used in smart order routing (SOR) to quantify fill probabilities, adverse selection and rebate/fee costs [Bacidore et al., 2010]. Tactical and SOR layers are entangled and an execution algorithm can have many non-linear dependencies within itself. Thus, an improvement in tactical and SOR levels doesn't necessarily translate linearly into better performance vs. benchmarks like IS or VWAP.  Order level TCA was studied in a different approach in [Rashkovich and Verma, 2012].   In this paper, we  study and calibrate models using placement level data.

IS benchmark is a primary TCA benchmark. IS measures the difference between arrival price and the average execution price. IS, a point-in-time benchmark, quantifies the actual alpha lost due to being unable to transact freely at the moment the decision to trade is made. For a small sample of discretionary trades, the benchmark can become biased by the cognitive decision-making and behavioral biases of a portfolio manager. 
IS  performance is a combination of volume and price timing, spread capture, adverse selection, short-term impact and reversion, venue selection and HFT anti-gaming logic. The claim is that implementation shortfall cannot be gamed and account for all transaction costs. Unfortunately, the IS benchmark is noisy and polluted by externalities (e.g., price momentum or behavior biases), accordingly, it often does not lead to actionable conclusions or understanding of what is the source of the slippage. We need other benchmarks to support the IS and increase chances that the IS out-performance is attributable to the broker's skills but not momentum externalities.  

Average price benchmarks like VWAP or PWP  provide a tradeable price with price momentum influence removed. In general, an average price benchmark implies not the best execution but an average one. 

The VWAP benchmark is often seen by traders as a fair price over a given time horizon, accounting for both volume and price. The VWAP benchmark is an average, not a point-in-time benchmark, giving it greater consistency and stability. It is also transparent and
easily understood by all market participants. Exposure to market risk is controlled by the duration of the execution. Good VWAP performance requires trading consistently at the  market rate throughout the order duration. Mathematically, trading with the VWAP benchmark leads to minimization of market impact.  Unfortunately, the real world is often not described by simple market impact  models;  multiple evidences from the industry show that VWAP executions  are expensive if  measured vs. the arrival price [Kingsley and Kan, 2016]. The VWAP benchmark can be gamed by aggressively front-loading the order and benefiting from excessive market impact at the beginning of the trade. Predictable schedule makes large VWAP orders easy prey for front-running by predatory high-frequency traders. Note that combining VWAP and IS better mitigates these scenarios than when using the VWAP benchmark alone.
As every trader knows, a perfect algorithm  executes aggressively at favorable prices, avoids paying high prices with a fair-value model, adjusts participation rate to the current price momentum,  crosses the spread to take advantage of available lit liquidity, seeks for a block crossing and randomizes schedule to avoid front-running by HFTs. Following a VWAP benchmark doesn't produce any incentives for these type of actions.

The PWP benchmark determines the volume-weighted average price of a stock from order start time with a given volume participation rate. PWP complements the traditional VWAP and IS benchmarks. 
In this study, we take aggressive  PWP 20 with a 20 percent participation rate.  It is believed that aggressive PWP 20 is a good proxy for a trader's skills in dealing with price momentum. Unfortunately, targeting  PWP 20 alone creates incentives to trade aggressively with a 20 percent participation rate,  leading to excessive costs.  
Trading aggressively in the beginning of a trade may match a PWP 20 benchmark but at the cost of an accumulated market impact that will stretch performance against arrival price and reversion. The variance of the benchmark has an unusual inverted U-shape with minimum at the target participation rate; in addition, there are industry reports that the mean of the benchmark may display non-monotonous behavior as a function of participation rate. 
In summary, minimizing PWP is not equivalent to minimizing trading costs.

The price reversion after the last fill is a proxy for an excessive pressure on the market. 
Obviously, the reversion is higher for poorly executed trades. The  reversion benchmark cannot be gamed and thus encourages the best trading practices.  

Based on the above, IS and reversion benchmarks are primary benchmarks for minimization of trading costs and best execution practice, with VWAP and PWP benchmarks being secondary ones. 


\section{Multiple Benchmarks TCA}

The industry standard is to use the VWAP benchmark for passive executions, but the IS benchmark for aggressive ones.  To quantify the performance of an arbitrary execution algorithm,  a TCA model should be viewed as a weighted combination of different benchmarks. 
Taking into account the multiple benchmarks helps to put an algorithm into a "cage" where any manipulation with benchmarks' distribution (e.g., having different rate of execution for an  in-the-money and an out-of-the-money position) cannot improve the final performance score calculated as a weighted sum of benchmarks.  For example, the simultaneous use of IS and price reversion benchmarks allows better measurement of the quality of the IS algorithm with a smaller data sample size.
In theory, we would like to have uncorrelated or anti-correlated benchmarks to cancel the noise, but it is not always possible in practice. Empirical Pearson correlation coefficients of the  benchmarks are shown in Appendix A. 

TCA with multiple benchmarks is  a good approach for evaluating  executions from different angles in order to obtain a more comprehensive view of a trading algorithm's  performance. Having said that, we believe that taking a weighted sum of the z-scores of point-in-time, average price benchmarks and reversion is the way to separate luck from skill and assign a fair performance score to an algorithm.

\section{Mathematical Formulation of TCA}
Mathematically,  the problem of an execution algorithm TCA is to estimate the conditional first moment of a TCA benchmark:
\be
E[y|x]=g(x,\theta)
\ee
here y is the performance against a TCA benchmark (i.e., IS, VWAP, PWP20, or Reversion), x is order or stock characteristics and g is an impact model function (unknown in general case) and $\theta$ is the parameters of the impact function.  The execution data consists of $N$ observations of $ (Y_N , X_N ) = {(y_1, x_1),(y_2, x_2), . . . ,(y_N , x_N )}$ where
$y_i \in R$ is a corresponding TCA benchmark variable  and $x_i \in R^d$  are $d$ covariates (order and stock characteristics). 
In this formalism, the square-root model of market impact has the form $E[IS|\sigma_D,Size, ADV] \sim \sigma_D \sqrt{\frac{Size}{ADV}}$. Here, IS is the implementation shortfall observations, $\sigma_D$ is the daily volatility, Size is the size of the order and $ADV$ is the average daily volume. 

In general, both the function $g$ and parameters $\theta$ are unknown. In the parametric approach, the form of the function $g$ is fixed and parameters $\theta$ are defined by calibration of the model using historical data. In the non-parametric approach, one can approximate the impact function $g$  by Gaussian processes or a neural network.   Since the best execution guidelines assume model transparency and interpretability, we take a  parametric approach with the primary objective of constructing a regression model to consistently estimate the parameters of interest $\theta$.

Our estimates represent a trading cost of an uninformed (on the time scale of the duration of order) trader.

\section{Econometrics of TCA Benchmarks}
In this section we discuss econometric properties of TCA benchmarks: heteroscedasticity, fat tails and non-Gaussianity, and skewness.  The benchmark distributions are shown in Appendix B.

\subsection*{Heteroscedasticity}

One of the key assumptions of regression is that the variance of the errors is constant across observations. Regression disturbances whose variances are not constant across observations are heteroscedastic. 
For example, the variance of  implementation shortfall of trades executed within two minutes  and  trades executed within two hours are different since the price volatility scales as the square root of duration of the execution. 
The same principle applies to VWAP, PWP and reversion variances as they all have  dependence on stock or order characteristics. 

For a nonlinear model, the standard methods (OLS) give biased estimates of regression coefficients if heteroscedasticity is not explicitly taken into account. 

\subsection*{Fat Tails and Non-Gaussianity}

All benchmark distributions have fat tails and pronouncedly non-Gaussian (peaky) shape. Fat tails make the convergence of the sample mean to the true mean slow. To visualize the painfully slow rate of convergence of a sample mean, we  applied  bucketing in order size and  participation rate for the IS benchmark. We  observe that the mean implementation shortfall is small relative to the variance. 
The error terms are significant and there are a number of inconsistent  results even when a  large number of  trades are averaged. Low signal-to-noise ratio makes  the process of building an impact model challenging.

The shape of distribution affects the way we estimate the expected value of a random variable given a finite number of observations. For example,  maximum likelihood estimator (MLE)  of the mean of  Gaussian random variables is given by a simple (arithmetic) mean. The MLE of the mean of  Laplace random variables is given by a median. 
In this paper, we use the full Bayesian approach, which supersedes the MLE approach. 

We attribute the non-Gaussian shape of benchmarks to the same nature of price and volume dynamics of a real market.

\subsection*{Skewness}

The benchmark observations are not spread symmetrically around an average value but, instead, have a skewed distribution.  Methods that assume a symmetric distribution cannot directly be applied to skewed data. Typically,  ignoring skewness of the distributions will cause a model to understate the risk (variance and tail risk) of variables with high skewness [Fleming, 2007].

When a benchmark distribution is modeled by the asymmetric Laplace distribution (which we discuss latter), the expected value of benchmark $y$ is given by:
\be
E[y]=\mu+\sigma*(\frac{1}{\kappa}-\kappa)
\ee
It is important to note that the expected value has a non-zero correction proportional to the scale parameter. It is typical for TCA benchmarks that the scale to location ratio is in the two to ten range that makes skewness correction significant even for small skewness parameter $\kappa$. For example, taking $\mu=-5$ bps, $\sigma=30$ bps and $\kappa=1.1$ the expected value 
\be
E[y]=-5-30*0.19=-5-5.7=-10.7\,\, bps.
\ee 
We see that even small skewness ($\kappa=1.1$) of the distribution doubles a naive estimate of the mean ($\mu=-5$ bps) due to large scale factor $\sigma$.  

The variance of the  expected value depends on location $\mu$, scale $\sigma$ and skewness $k$ parameters and  is given by:
\be
\Sigma^2=E[y^2]-E[y]^2=\sigma^2 \frac{1+\kappa^4}{\kappa^2}
\ee
for $k=1.1$ the correction factor is ${\frac {1+\kappa ^{4}}{\kappa ^{2}}}=2.03$.

We attribute some of the skewness of benchmarks to the maneuvering of  trading algorithms like slowing executions when underperforming against the benchmark or accelerating executions when outperforming against the benchmark.  While this strategy works well in a mean-reverting regime, it fails in momentum-driven markets. 

\section{Generic TCA Regression Model}

In this section we quantify the dependence of the expected value $E[y]$  of a TCA benchmark y on stock characteristics (spread and volatility) and order parameters (Size/ADV and participation rate). The $\pi(E[y])$ distribution of the  posterior of the expected value $E[y]$ is the ultimate quantity of our interest.


The data shows that the asymmetric Laplace distribution(ALD) [Wikipedia: Asymmetric Laplace distribution] $ALD(\mu,\sigma,k)$ is a plausible assumption for distribution of the benchmarks. The probability density function (pdf) of ALD $P_{ALD}$ is given by: 
\be
{\displaystyle P_{ALD}(y;\mu,\sigma ,\kappa )=\left({\frac {1}{\sigma (\kappa +1/\kappa) }}\right)\,e^{-\frac{(y-\mu)}{\sigma} \,s\kappa ^{s}}}
\ee
here $s=sgn(y-\mu)$, $\kappa$ is asymmetry parameter and scale parameter $\sigma$. The mean E[y] of random variable $y$ is given by:
\be
E[y]=\mu+\sigma*(\frac{1}{\kappa}-\kappa)
\label{eq:expectation}
\ee

Assuming a benchmark $y$ is  distributed according to $ALD$, we use the following Bayesian regression  model: 
\be
P(y |\mathbf {X} ,{\boldsymbol {\beta  }},{\boldsymbol {\gamma}},{\boldsymbol {\alpha}}) \sim ALD(\mu,\sigma,\kappa)
\label{eq:regression}
\ee
to estimate location parameter  $\mu$, scale parameter $\sigma$ and asymmetry parameter $\kappa$  as a function of  independent variables $\mathbf {X}$:
\be
X_{1}={Size/ADV}, \,\, X_{2}={\rho}, \,\, X_{3}={\sigma_{D}}, \,\, X_{4}={S}, 
\ee
here $Size$  is the size of the order (in shares), ADV is 30-days historical average daily volume, $\rho$  is participation rate (in percent), $\sigma_{D}$  is 30-days annualized historical volatility (in percent), and $S$ is the 5-day historical average spread $S$ of a stock (in bps).  
The dependent variable $y$ is a TCA benchmark: IS, VWAP, PWP or Reversion. 

Cost models often have a multiplicative form that dictates an exponential parametrization of the independent factors $X_{1,2,3,4}$. The location parameter $\mu$ is given by:
\be
\mu=-\exp{\mu_{ln}}; \,\,\,\mu_{ln}=\beta_0+\beta_1\ln(X_1)+\beta_2\ln(X_2)+\beta_3\ln(X_3)+\beta_4\ln(X_4)
\label{eq:mu}
\ee
The heteroscedasticity  of the benchmark $y$ (except PWP 20)  is taken into account by parameterizing the scale parameter $\sigma$ as follows: 
\be
\sigma=\exp{\sigma_{ln}}; \,\,\, \sigma_{ln}=\gamma_0+\gamma_1 \ln(X_1)+\gamma_2 \ln(X_2)+\gamma_3 \ln(X_3)+\gamma_4 \ln(X_4)                    
\label{eq:sigma1}
\ee

The heteroscedasticity of the PWP 20 benchmark is a special case as the scale parameter $\sigma$  vanishes around 20 percent participation rate and we have to add an additional component $\gamma_5$ to take this into account: 
\be
\sigma=\exp{\sigma_{ln}}; \,\,\, \sigma_{ln}=\gamma_0+\gamma_1 \ln(X_1)+\gamma_2 \ln(X_2)+\gamma_3 \ln(X_3)+\gamma_4 \ln(X_4) +\gamma_5 \ln(|X_2-20|+\gamma_6)                    
\label{eq:sigma2}
\ee


The functional form of skewness was chosen to make two terms in equation  ($\ref{eq:expectation}$) to have the same functional form. Thus $r=(\kappa-\frac{1}{\kappa})$ should have the same multiplicative form as location $\mu$ and scale $\sigma$ parameters.
We parametrize $r$ as:
\be
r=\exp{\alpha_{ln}};  \,\,\, \alpha_{ln}=\alpha_0+\alpha_1\ln(X_1)+\alpha_2\ln(X_2)
\label{eq:r}
\ee
This ansatz is justified as for all empirical data we have $\kappa\ge1$ ($\kappa=1$ corresponds to zero skewness). Our numerical experiments show that the order parameters $X_1$ and $X_2$ are the major factors that influence the skewness of the distribution.
The skewness parameter $\kappa$ can be expressed as a function of $r$: 
\be
\kappa=\frac{1}{2} \left(r+\sqrt{4+r^2} \right)           
\ee 

To calibrate the generic TCA model, we aggregate execution data from all algorithms per region. We discuss the detailed procedure of calibration in Appendix C and Appendix D.  The posterior predictive distributions are shown in Appendix F. 

 
A similar non-Bayesian form of this regression with a symmetric Gaussian error term was  investigated in [Engel et al., 2008]:
\be
y=-\exp{(\sum_{k}  \beta_k \ln{X_{k}})}+\exp{(\sum_{k} \gamma_k \ln{X_{k}} )}\times \epsilon. 
\ee
where $\epsilon$ is a noise term.  The noise term $\epsilon$ was taken to be a Gaussian variable and different independent variables $X_k$ were used.

In summary, the benchmark distributions are fat-tailed, skewed and heteroscedastic. All these features violate standard regression assumptions and have to be addressed explicitly. Bayesian inference is the right tool here as it requires the explicit  statement of  all underlying assumptions.


\section{Algorithm-Specific TCA Regression Model}
Often a trader is interested in performance estimation of a particular broker execution algorithm. In this case, the main problem is a limited sample size.  Bayesian modeling allows us to overcome this problem by using  hierarchical partially pooled models.

The modeling is done in two stages. First, we build the pooled regression model described in the previous section where execution observations  from algorithms are aggregated, algorithm labels are discarded  and the generic TCA  model is fitted on this dataset.  
This allows for calculation of the posterior distribution of regression coefficients $\alpha,\beta$ and $\gamma$ in Equations (\ref{eq:mu}),(\ref{eq:sigma1}), and (\ref{eq:r}). The posterior distribution has two competing contributions: a single contribution from the  prior and likelihood contribution from each observation which increases linearly with the number of observations. Due to the large number of observations in the aggregated sample, the posterior is dominated by the likelihood term and is robust to the parameters of the prior. 

Second, we use a partially pooled version of regression (\ref{eq:regression}) to build the algorithm-specific TCA regression. The coefficients  $\alpha$ and $\gamma$ are broker-specific (unpooled). The  heteroscedasticity coefficients $\beta$ of the scale parameter $\sigma$ are identical (pooled) for all algorithms as they characterize the benchmark but not an algorithm.   
The posterior of the generic model from the first stage is used as a prior for the algorithm-specific TCA regression. 

Markov Chain Monte Carlo (MCMC) method is used to obtain samples from the posterior distribution of regression coefficients $\alpha$, $\beta$ and  $\gamma$. It allows for calculation of  the distribution of the expected value of a TCA benchmark $\pi(E[y])$ using the formula for the expected value of the asymmetric Laplace distribution  ($\ref{eq:expectation}$)  given order and stock parameters.

Note that if a particular broker has only a small number of executions, the approach still holds. In this case, the posterior distribution of a TCA  benchmark $E[y]$ is dominated by the prior that is given by the generic TCA model. 
It means that the cost will be an average one  but never the best one. To have the best trading cost, the likelihood contribution of a particular broker should overcome the gravity of the prior of the generic TCA model by having superior execution quality and enough of them in the data sample.


The use of hierarchical models allows diffusion of information from a large aggregated sample to a smaller data sample of a particular algorithm. The two-stage procedure ensures that ranking of algorithms by the trading cost is robust even if the number of observations is small for a particular algorithm.  

\section{Ranking of Trading Algorithms}
We rank algorithms in a two-stage process. 

First, given the order ($X_1$ and $X_2$) and stock ($X_3$ and $X_4$)  parameters, we calculate the relevance score $R$ for each algorithm.   The objective here is to calculate distance between order and stock characteristics and their corresponding historical distributions for given algorithms. The Mahalanobis distance $d_M$ [Wikipedia: Mahalanobis distance], which is a measure of the distance between a point $K$ and a distribution $D$, provides a convenient formalism for it.  
We calculate distance $d_M^{order}$  between  $ \ln X_1$ and $\ln X_2$ and its joint two-dimensional historical distributions for each algorithm and distance $d_M^{stock}$  between  $\ln X_3$ and $\ln X_4$ and its two-dimensional historical distributions for each algorithm. The distances are converted to bounded z-scores $z_{range}$ ($z_{range}=100*\tanh(z)$ of the original z-score) and the weighted sum is taken to get the relevance score $R$:
\be
R=w_r^{order} z_{range} (d_M^{order})+w_r^{order} z_{range} (d_M^{stock})
\ee
here, the default values of $w_r^{order}=\frac{2}{3}$ and $w_r^{stock}=\frac{1}{3}$. Note that the two-dimensional historical distribution of order parameters can be multi-modal due to the discrete nature of aggressiveness for some algorithms. In this case, the distance  $d_M^{order}$  is the distance to the nearest cluster. Next,  we keep only 20 percent of the original algorithms ranked by relevance. This ensures  that the selected set of algorithms was actively used in the past based on historical distribution of  order and stock parameters.  It also provides an extra guarantee that parametric TCA model output is calculated close to the center of mass of training data points and the TCA regression model was used  in the interpolation but not in the extrapolation regime.  

Second, we calculate the performance score $P$. The performance score $P$ is defined as a weighted sum (the sum of weights is normalized to one) of  bounded z-scores ($z_{range}= 100*\tanh(z)$ of the original z-score) of each relevant benchmark. Since human traders are more comfortable with a deterministic setup, we take the mean $E[y]$ of algorithm's posterior distribution of benchmark $y$  as the input to the z-score. Alternatively, to take into account the specific properties of the trading cost distribution,  a probabilistic ranking procedure  can be used.
In this case, the trading cost is not deterministic but is given by a realization of a random variable sampled from algorithm  ($A$) specific trading cost distribution $\pi_{A} (E[y])$. For each routing decision (algo wheel), the best broker is selected based on the sampled trading cost.  The posterior distribution of the IS and VWAP means are shown in Appendix G.
The weights encode a trader's subjective view on  importance of benchmarks that should serve to preserve alpha in a portfolio manager strategy.  The industry standard is to measure large orders with low participation rate with large VWAP weight  $w_{VWAP}$  while aggressive orders have large weight  of IS benchmark $w_{IS}$. 

As a final step, we rank algorithms based on a total score $T$ which is defined as a weighted sum of the relevance $R$ and performance scores $P$: 
\be
T=w_r R+w_p P
\ee
here, the default values of $w_r=0.3$ and $w_p=0.7$. We believe that balancing relevance and performance represents a holistic view on ranking trading algorithms and prevents the appearance of pathological edge cases via the  self-correction mechanism.

\section{Acknowledgments}
The author would like to thank Andrei Iogansen, Tito Ingargiola, Kapil Phadnis, Vlad Rashkovich and Vasilisa Markov for helpful discussions and suggestions.
\section{Conclusion}
In this paper we presented a formulation of the transaction cost analysis  in the Bayesian framework. Our formulation allows us to effectively calculate the 
expected value of trading benchmarks with only a finite sample. We also discussed the application of the method to ranking of broker execution algorithms. 


\clearpage
\section*{Appendix A: Correlation Between Benchmarks}
Pearson correlation matrices of  TCA  benchmarks using four participation rate buckets are shown in Table 2.
\FloatBarrier

\begin{table}[htbp]
\caption*{Participation rate range: $1\text{-}7$ percent } 
\centering 
\begin{tabular}{|l|r|r|r|r|}\hline
           &         IS &         VWAP&         PWP20 &         Rev5m \\
\hline
IS               &   1 &  -0.01 &  0.79 &  0.04 \\
\hline
VWAP        &  -0.01 &  1 & 0.07 &  0.03  \\ 
\hline
PWP20      &   0.79 &  0.07 & 1 &  0.04 \\ 
\hline
Rev5m    &   0.04 & 0.03 &  0.04 &  1 \\ 
\hline
\end{tabular}
\\
\caption*{Participation rate range: $7\text{-}15$ percent} 
\centering 
\begin{tabular}{|l|r|r|r|r|}\hline
           &         IS &         VWAP&         PWP20 &         Rev5m \\
\hline
IS               &   1 &  0.12 &  0.63 &  0.07 \\
\hline
VWAP        &  0.12 &  1 & 0.38 &  0.09  \\ 
\hline
PWP20      &   0.63 &  0.38 & 1 &  0.09 \\ 
\hline
Rev5m    &   0.07 & 0.09 &  0.09 &  1 \\ 
\hline
\end{tabular}
\end{table}
\begin{table}[htbp]
\caption*{Participation rate range: $15\text{-}25$ percent} 
\centering 
\begin{tabular}{|l|r|r|r|r|}\hline
           &         IS &         VWAP&         PWP20 &         Rev5m \\
\hline
IS               &   1 &  0.23 &  0.35 &  0.11 \\
\hline
VWAP        &  0.23 &  1 & 0.83 &  0.05 \\ 
\hline
PWP20      &   0.35 &  0.83 & 1 &  0.10 \\ 
\hline
Rev5m    &   0.11 & 0.05 &  0.10 &  1 \\ 
\hline
\end{tabular}
\caption*{Participation rate range: $25\text{-}40$ percent}
\centering 
\begin{tabular}{|l|r|r|r|r|}\hline
           &         IS &         VWAP&         PWP20 &         Rev5m \\
\hline
IS               &   1 &  0.40 &  0.10 &  0.27 \\
\hline
VWAP        &  0.40 &  1 & 0.43 &  0.15  \\ 
\hline
PWP20      &   0.10 &  0.43 & 1 &  0.27 \\ 
\hline
Rev5m    &   0.27 & 0.15 &  0.27 &  1 \\ 
\hline
\end{tabular}
\caption*{Table 2. Pearson correlation matrices using four participation rate buckets.}
\end{table}
\FloatBarrier

\newpage 
\section*{Appendix B: Distribution of Trading Benchmarks}
The distribution of TCA benchmarks in the U.S. is shown in Figure 1. The distributions have pronouncedly non-Gaussian shape: shape is peaky, tails are fat and the data is skewed to worse executions. 
We show distribution of IS, VWAP, PWP 20 and 5 min Reversion benchmarks for pooled (aggregated) data and for the VWAP, POV and IS algorithms of a major broker in the U.S.
Note that the VWAP algorithm (low participation rate)  benchmark distributions are almost symmetric, while the IS and POV algorithms (high participation rate) benchmark distributions are skewed.

\begin{figure}[ht]
  \captionsetup[subfigure]{labelformat=empty}
  \hspace*{\fill}%
  \subcaptionbox{All Algorithms}{\includegraphics[width=2.7in]{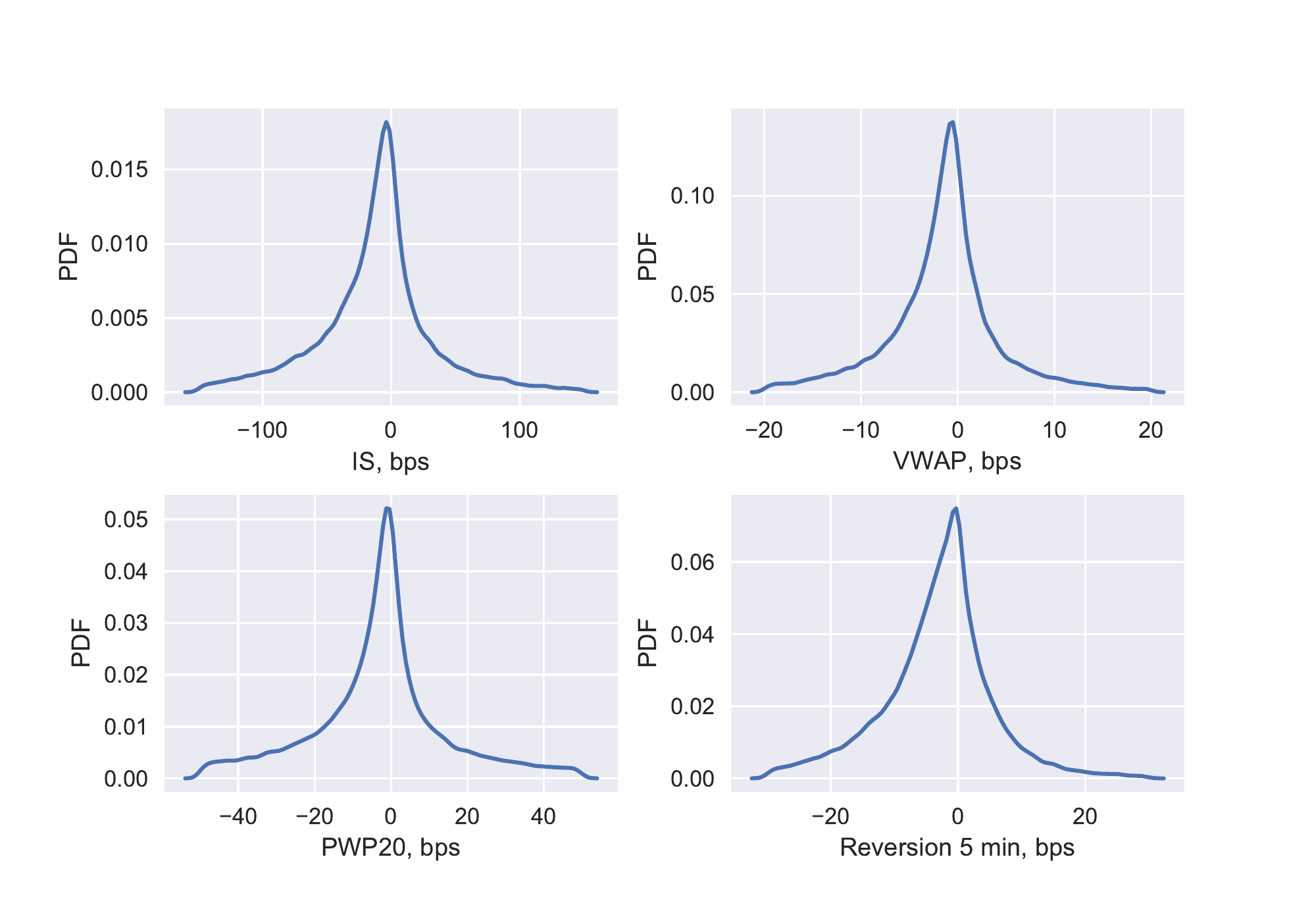}}\hspace{1em}%
  \subcaptionbox{A VWAP algorithm}{\includegraphics[width=2.7in]{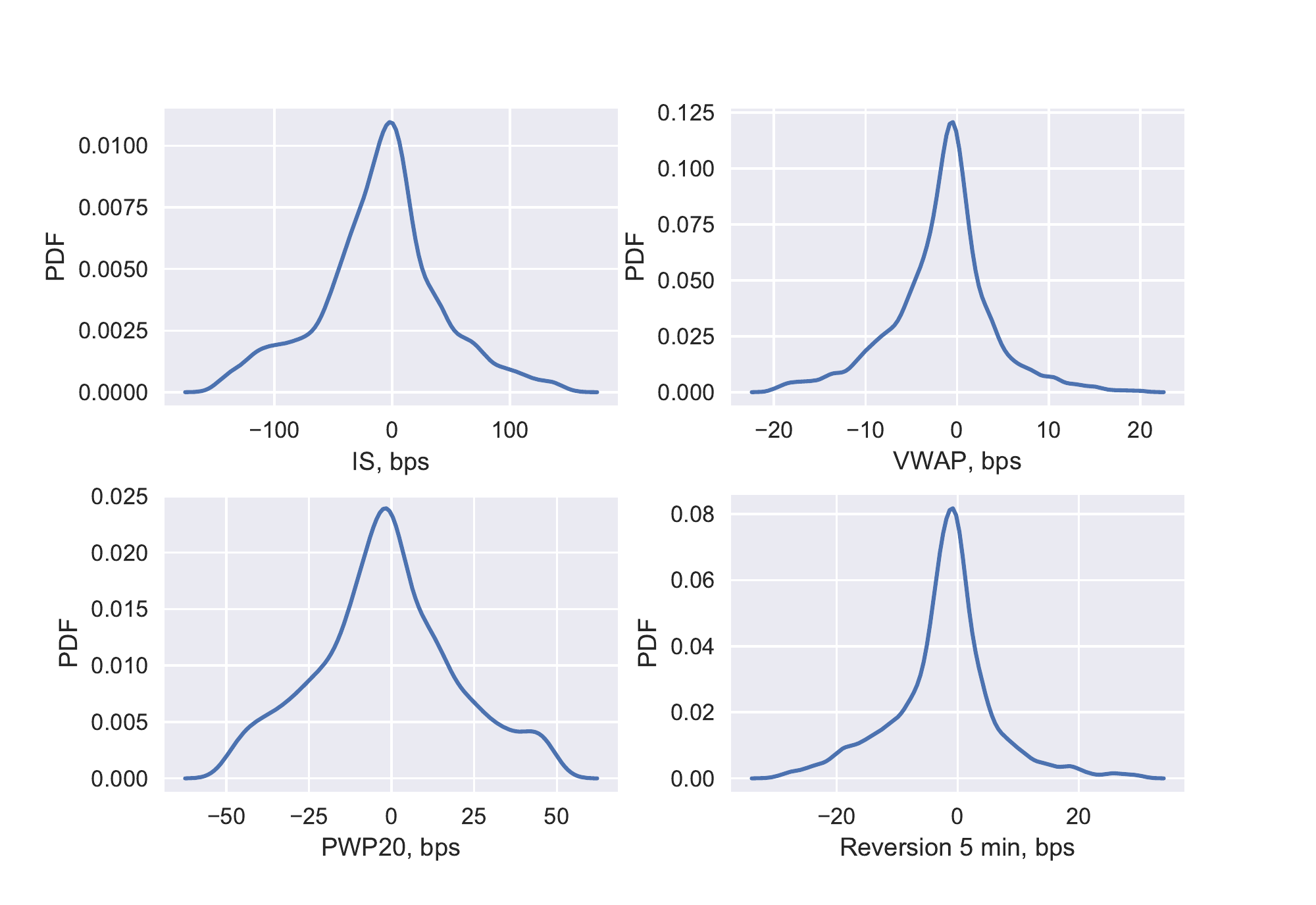}}%
  \hspace*{\fill}%
\end{figure}

\begin{figure}[ht]
  \setcounter{figure}{0}
  \captionsetup[subfigure]{labelformat=empty}
  \hspace*{\fill}%
  \subcaptionbox{A POV algorithm}{\includegraphics[width=2.7in]{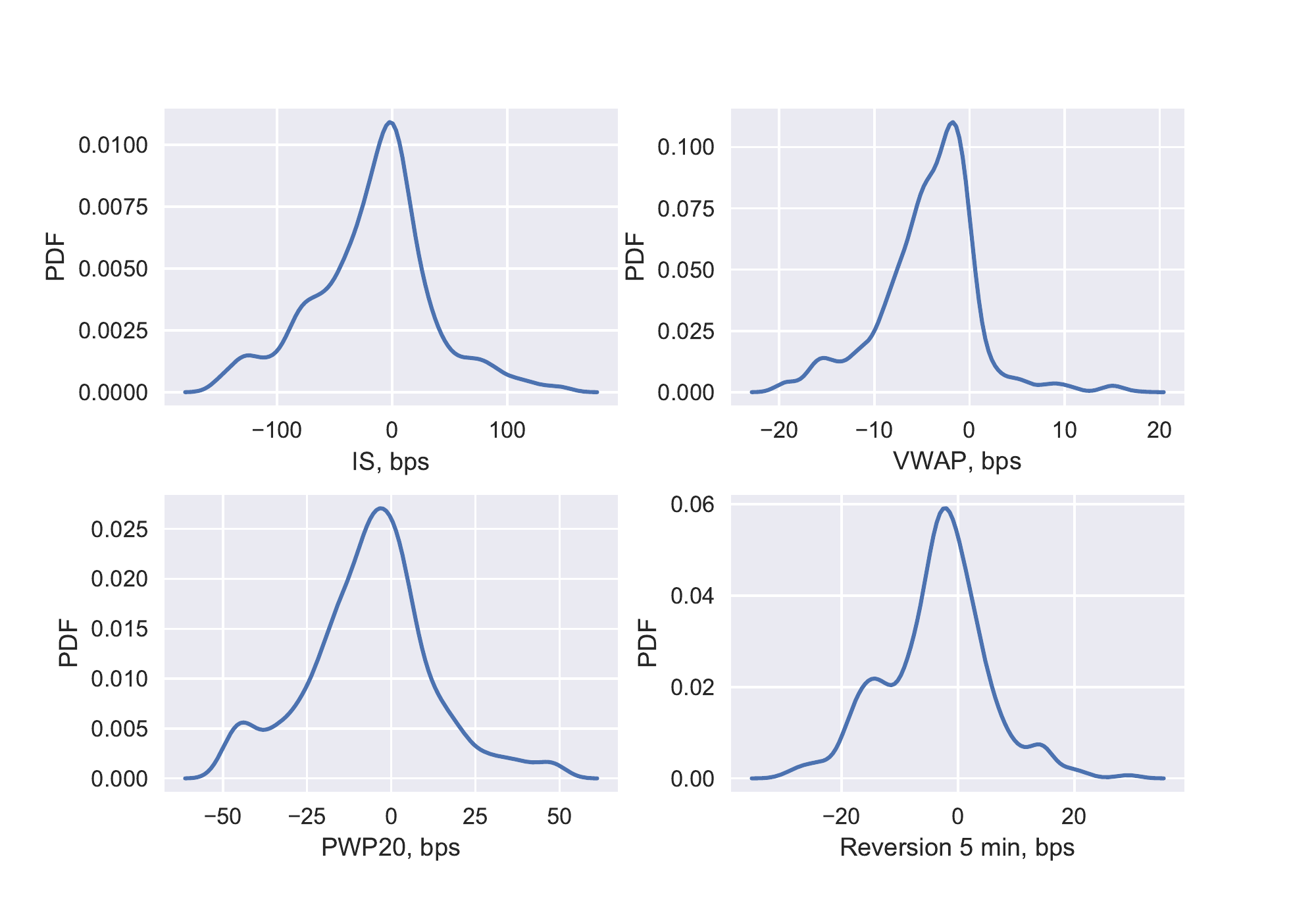}}\hspace{1em}%
  \subcaptionbox{An IS algorithm}{\includegraphics[width=2.7in]{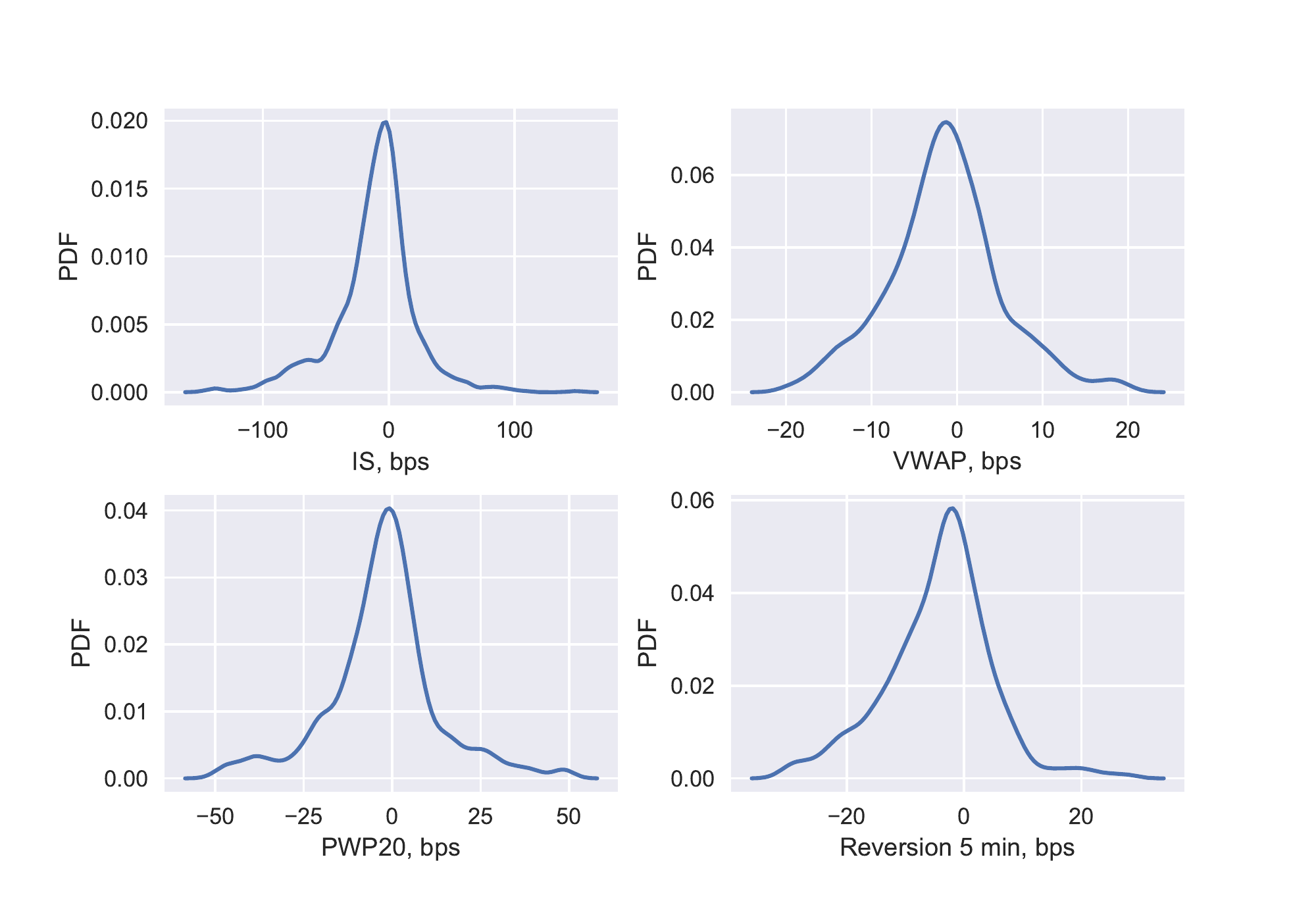}}%
  \hspace*{\fill}%
  \label{fig:tca_benchmarks}
  \caption{Empirical probability distributions of TCA benchmarks for all available data and  VWAP, POV and IS algorithms of a major broker in the U.S.}
\end{figure}


\section*{Appendix C: The Regression Model Formulation}
The prior for the generic pooled regression coefficients were chosen as follows.  Coefficients  $\beta$ for the location parameter  $\mu$  are normally distributed $N(\hat \mu, \hat \sigma)$ with mean $\hat \mu$ and standard deviation $\hat \sigma$. For $\beta_0$, we have a wide prior; and for $\beta_{1,2,3,4}$, we choose a concave form: 
\be
\beta_0\sim  N(0,2), \,\,\, \beta_{1,2,3,4}\sim  N(0.5,0.5) 
\ee
For $\gamma_0$, we have a wide prior;  for $\gamma_{1,2,3,4,5}$, we choose a concave form:  
\be
\gamma_0\sim  N(0,2), \,\,\, \gamma_{1,2,3,4,5}\sim  N(0.5,0.5),  \,\,\ \gamma_{6}\sim  N(1,1)  
\ee
For $\gamma_{6}$, we use the normal distribution which is bounded  by zero on the left side.

Given the sensitivity of the expected value to the skewness parameter $\kappa$,  we choose $\alpha_{0}$ to be a negative number and $\alpha_{1,2}$ to be centered around zero so that we start with zero skewness assumption and only adjust skewness if data so dictates: 
\be
\alpha_0\sim  N(-5,2),\,\,\,  \alpha_{1,2}\sim  N(0,0.5)   
\ee

The algorithm-specific hierarchical regression model takes the mean $\mu^{post}_i$ and standard deviation  $\sigma^{post}_i$ of  the posterior distribution  of  $\beta^{post}_i$ and $\gamma^{post}_i$ and $\alpha^{post}_i$ as a prior for algorithm $A$  specific coefficients $\beta^A_i$ and $\gamma_i$ and $\alpha^A_i$. 
The regression coefficients $\beta^A_i$ and $\gamma_i$ and $\alpha^A_i$ priors are normally distributed. 

The coefficients  $\beta$  of the location parameter $\mu$  are unpolled  (different for each algorithm) and for algorithm $A$ are given by:
\be
\beta^A_i \sim  N\left(\mu({\beta^{post}_i}),\sigma({\beta^{post}_i})\right)
\ee
The coefficients  $\gamma$  of the scale parameter $\sigma$ are pooled  (identical for all algorithms) as they represent the property of a benchmark but not the algorithm. 
\be
\gamma_i \sim  N\left(\mu({\gamma^{post}_i}),\sigma({\gamma^{post}_i})\right)
\ee
The coefficients  $\alpha$ of the skewness  parameter $\kappa$ are unpolled  (different for each algorithm) and for algorithm $A$ are given by:
\be
\alpha^A_i \sim  N\left(\mu({\alpha^{post}_i}),\sigma({\alpha^{post}_i})\right)
\ee

According to the Bayes theorem, the posterior distribution of regression coefficients $w \in [\beta,\gamma,\alpha]$  given the observations $y$ is: 
\be
P(w |y)\propto P(y|w) P(w)
\ee
The prior $P(w)$ is defined above and likelihood  $P(y|w)$  is given by the asymmetric Laplace distribution (\ref{eq:regression}). Alternatively, one can use the asymmetric generalized normal distribution or asymmetric t-distribution for the likelihood. They may provide better fit due to an additional shape parameter, but we have found calibration to be less stable for our problem.  
Samples from posterior distributions $P(w |y)$ are obtained using Markov Chain Monte Carlo method [Brooks S. et al, 2011].



\section*{Appendix D: Calibration of the Generic Model}

To demonstrate the practicality of our approach, we calibrated the generic regression model on a subset of Bloomberg EMSX data. The first dataset has  67 thousand observations in the U.S. (U.S. model); the second dataset has 61 thousand observations from the UK, Germany, France, Italy, Spain and Switzerland (EU model).  The data runs from March 21, 2017 until September 21, 2018. We selected fully completed market (no limit price) day orders with duration of more than 5 minutes, maximum $Size/ADV$  20 percent, minimum $Size/ADV$ $0.1$ percent,   maximum participation rate 40 percent, and  minimum participation rate 1 percent. 

We  used the standard definition of trading benchmarks measured in basis points. Let us denote the average execution price as $\bar P$, $S$ is the sign of the trade ($S=1$ for a buy trade and $S=-1$ for a sell trade),  then the benchmarks are given by: 
\be
IS = \frac{Arrival Price – \bar P}{\bar P}* S*10000
\ee
\be
VWAP = \frac{P_{VWAP} – \bar P}{\bar P}* S*10000
\ee
\be
PWP20 = \frac{P_{PWP 20} – \bar P}{\bar P}* S*10000
\ee
For 5-minute reversion, we take a difference between last fill $P_{last fill}$ and VWAP price 5-minutes after the last fill $P_{VWAP}^{5 min}$.

\be
Rev5m= \frac{ P_{VWAP}^{5 min}-P_{last fill} }{P_{last fill}}* S*10000
\ee
Additionally, we only take observations that satisfy all of the following cutoffs ($|benchmark|<c$): $c_{IS}=500$ bps,  $c_{VWAP}=150$ bps, $c_{PWP20}=150$ bps and $c_{Rev5m}=200$ bps.

For calibration we used  PyMC3 [Salvatier J. et al., 2015] implementation of Metropolis-Hastings method with a large number of iterations  ($N_{iter}=500000$, $ N_{burn}=400000$, $N_{thining}=20$) and  No-U-Turn Sampler (NUTS) with ($N_{iter}=10000$, $N_{burn}=5000$) and got identical results (within a small numerical error). 
The mean $\mu_{benchmark}$ and standard deviation $\sigma_{benchmark}$ of the marginal distribution of regression coefficients of the model  calibrated with NUTS algorithm are shown in Table 1.

\begin{table}
\resizebox{\columnwidth}{!}
{
\begin{tabular}{|l|r|r|r|r|r|r|r|r|r|r|r|r|r|r|r|}\hline
{US} &         $\beta_0$ &         $\beta_1$ &         $\beta_2$ &         $\beta_3$ &         $\beta_4$ &         $\gamma_0$ &          $\gamma_1$ &         $\gamma_2$ &      $\gamma_3$ &    $\gamma_4$ &  $\gamma_5$ &   $\gamma_6$ &    $ \alpha_0$ & $ \alpha_1$ & $ \alpha_2$ \\
\hline
$\mu_{IS}$ &  0.89 &  0.46 &  0.09 &  0.83 &  0.16 &  3.76 &  0.43 & -0.45 &  0.64 &0.2& -& -& -3.4&-0.22 &0.49 \\ $\sigma_{IS}$ &      0.14 &  0.02 &  0.03 &  0.03 &  0.02 &  0.03 &  0.0 &  0.0 &  0.01 &0.01& -& -&  0.16&0.03&0.04 \\
\hline
$\mu_{VWAP}$  &  -1.12 &  0.08 & 0.0 &  0.01 &  0.85 &  0.84 &  0.18 & -0.33 &  0.43&0.36 & -& -& -5.1& -0.46 & 0.13  \\ $\sigma_{VWAP}$ &     0.1 &  0.01 &  0.02 &  0.02 &  0.02 &  0.03 &  0.00 &  0.0 &  0.01&0.01 & -& -&  0.43 &0.07& 0.08  \\
\hline
$\mu_{PWP20}$ &  -0.02 &  0.14 & -0.32 &  0.13 &  0.72 & -0.2 &  0.33 & -0.5 &  0.63&0.19 & 1.04&5.91&  -3.45&-0.28&0.24 \\ $\sigma_{PWP20}$ &   0.19 &  0.02 &  0.04 &  0.04 &  0.03 &  0.19 &  0.0 &  0.01 &  0.1&0.01 &0.05& 0.52&0.19 &  0.03&0.03\\
\hline
$\mu_{Rev5m}$ &    -2.93 & -0.09 &  0.47 &  0.17 &  0.71 & -0.45 & -0.02 &  0.07 &  0.53 &0.24& -& -&  -1.73&0.00&0.24 \\ $\sigma_{Rev5m}$ &    0.17 &  0.03 &  0.03 &  0.03 &  0.03 &  0.03 &0.0&  0.0 &  0.01 &0.01&  -& -& 0.15 &  0.03& 0.03 \\
\hline
{EU} &         $\beta_0$ &       $  \beta_1$ &         $\beta_2$ &        $ \beta_3$ &        $ \beta_4 $&        $ \gamma_0$ &          $\gamma_1$ &          $\gamma_2$ &         $ \gamma_3$ &   $\gamma_4$ &     $\gamma_5$ &    $\gamma_6$ &      $ \alpha_0$ & $ \alpha_1$ & $ \alpha_2$ \\
\hline
$\mu_{IS}$ &    2.29 &  0.44 &  0.01 &  0.47 &  0.09 &  4.44 &  0.45 &-0.45& 0.47 &  0.13 &  -& -& -3.4&-0.12&0.57   \\$ \sigma_{IS}$ &     0.13 &  0.02&  0.02 &  0.02 &  0.02 &  0.04 &  0.00 &  0.00 &  0.01&0.01 & -& -&  0.19&0.03&0.04 \\
\hline
$\mu_{VWAP} $ & -0.76 &  0.17 &  0.04 &  0.25 &  0.33 &  2.27 &  0.31 & -0.41&  0.34 & 0.19& -& -& -6.04&-0.47&0.33\\ $\sigma_{VWAP}$ &      0.14 &  0.02 &  0.02 &  0.03 &  0.02 &  0.04&  0.00 &  0.00 &  0.01&0.0&-& -&   0.76&0.14&0.12  \\
\hline
$\mu_{PWP20} $& 1.34 &  0.29& -0.39 &  0.26 &  0.27 &  1.09 &  0.41 & -0.54 &  0.42 &  0.12&1&5.47& -3.23&-0.2&0.28 \\ $\sigma_{PWP20}$ &      0.22 &  0.02 &  0.04 &  0.05 &  0.03 &  0.19 &  0.00 &  0.01 &  0.01&0.01 & 0.05&0.5&   0.21&0.04&0.04\\
\hline
$\mu_{Rev5m}$ & -0.74 &  0.11 &  0.31 &  0.28 &  0.19 &  1.76 &  0.26 & -0.2 &  0.54 &0.2&  -& -& -2.18 &0.07&0.31\\ $\sigma_{Rev5m}$ &   0.19 &  0.02 &  0.04 &  0.04 &  0.02 &  0.04 &  0.00 &  0.01&  0.01&0.01&-& -& 0.25 &  0.04&0.08  \\
\hline
\end{tabular}
}
\caption*{Table 1. The coefficients of the Generic TCA Model.}
\end{table}
\newpage
The regression coefficients of the U.S. model have clear interpretation. The location parameter $\mu$ of the IS benchmark has roughly the square-root law behavior:  $\mu \sim \sigma_D^{0.99}\times (Size/ADV)^{0.46}$ (assuming $S \sim \sigma_D$). The scale parameter $\sigma$ scales approximately as the square root of the order duration $T\sim \frac{Size/ADV}{\rho}$:  $\sigma \sim \sigma_D^{0.84}\times T^{0.43}$ (assuming $S \sim \sigma_D$).
The location parameter $\mu$ of the VWAP benchmark is mainly a function of the  spread parameter $\mu\sim S^{0.85}$.  The dependence of the expected value of the PWP 20 benchmark on the participation rate $\rho$ is non-monotonous and is dictated by the $\gamma_5=1.04$  coefficient. The location parameter $\mu$ of the Reversion benchmark depends on the participation rate $\mu \sim \rho^{0.47}$ but is not affected by the $Size/ADV$ parameter. 
\clearpage
\section*{Appendix F: Model Validation}
The posterior predictive distributions of the generic model for the U.S. and EU are shown in Figures 2 and 3. 
\begin{figure}[ht!]
  \captionsetup[subfigure]{labelformat=empty}
  \hspace*{\fill}%
  \subcaptionbox{Figure 2: The posterior predictive distributions for the U.S. Generic Model.}{\includegraphics[width=6in,height=3in]{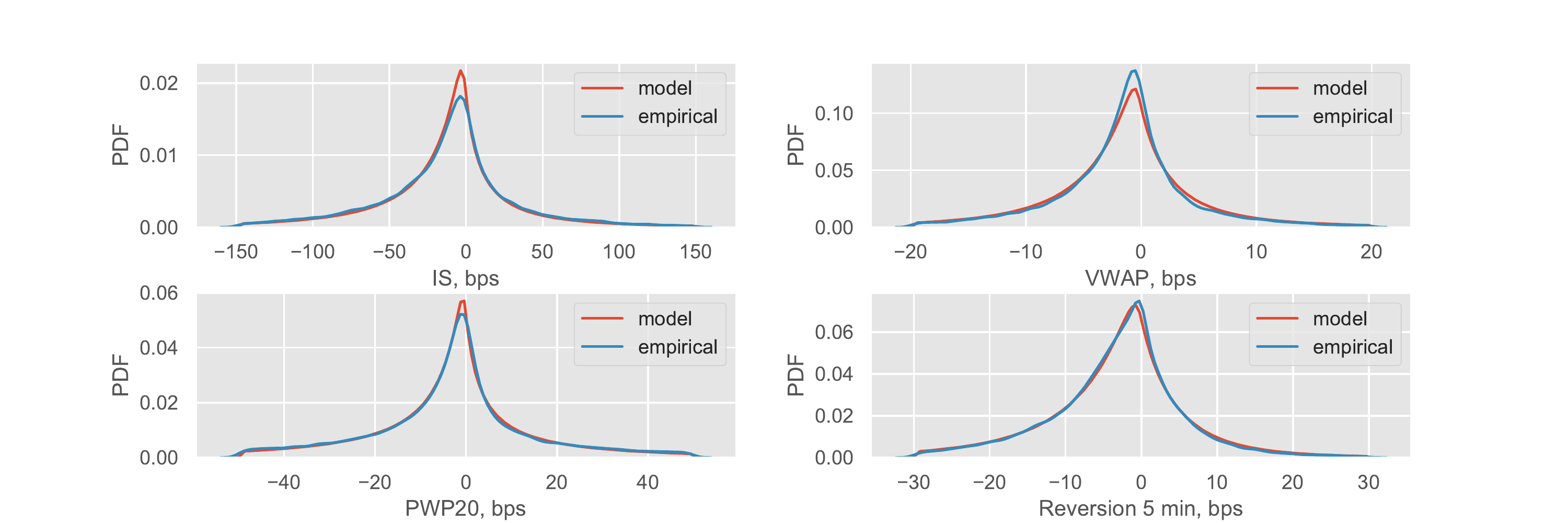}}\hspace{1em}%
  \subcaptionbox{Figure 3: The posterior predictive distributions for the EU Generic Model.}{\includegraphics[width=6in,height=3in]{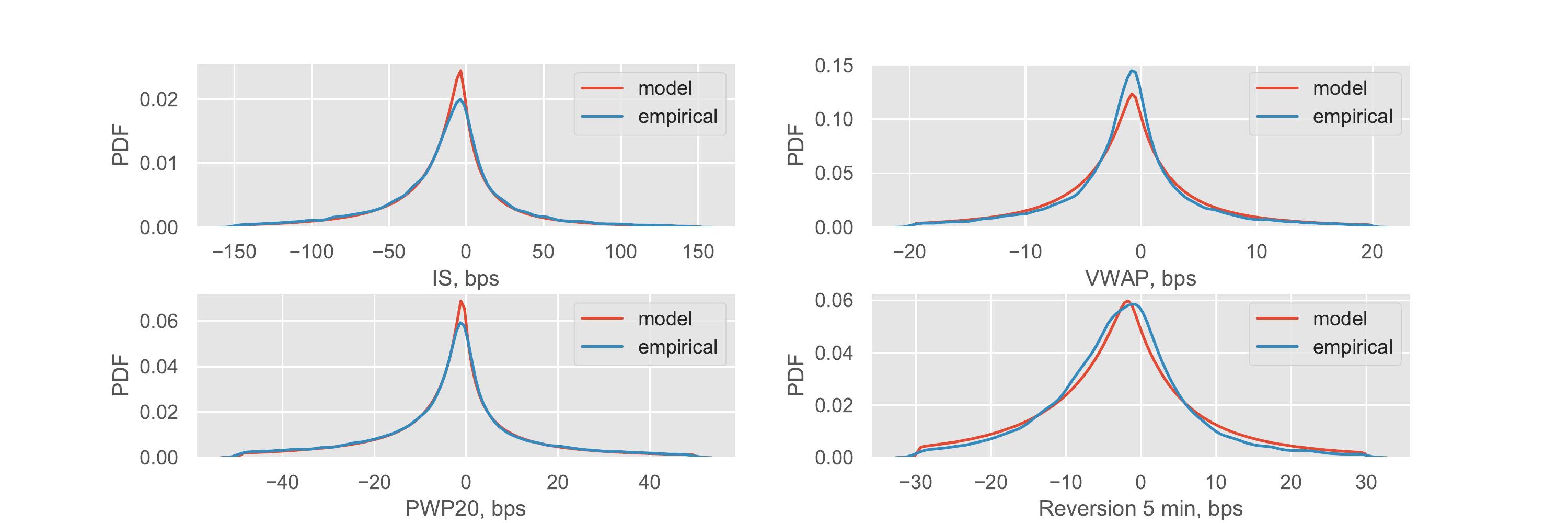}}\hspace{1em}%
  \hspace*{\fill}%
\end{figure}

\clearpage
\section*{Appendix G: The Posterior Distribution of the IS and VWAP Costs}
The posterior distribution of the expected value of the IS benchmark of an order with high participation rate (Size/ADV=0.01, participation rate=25 percent, annualized volatility 30 percent and spread 10 bps) for three IS strategies of major U.S. brokers are shown in Figures 4. 
The posterior distribution of the expected value of the VWAP benchmark of an order with low participation rate  (Size/ADV=0.03, participation rate=3 percent, annualized volatility 30 percent and spread 10 bps) for three VWAP strategies of the same major U.S.  brokers are shown in Figures 5. 
Numerical experiments show that  IS and reversion costs are good strategy discriminators while VWAP cost is a mediocre one. 
\begin{figure}[h]
  \captionsetup[subfigure]{labelformat=empty}
  \hspace*{\fill}%
  \subcaptionbox{Figure 4: The posterior distributions of the IS cost.} {\includegraphics[width=6in,height=2.5in]{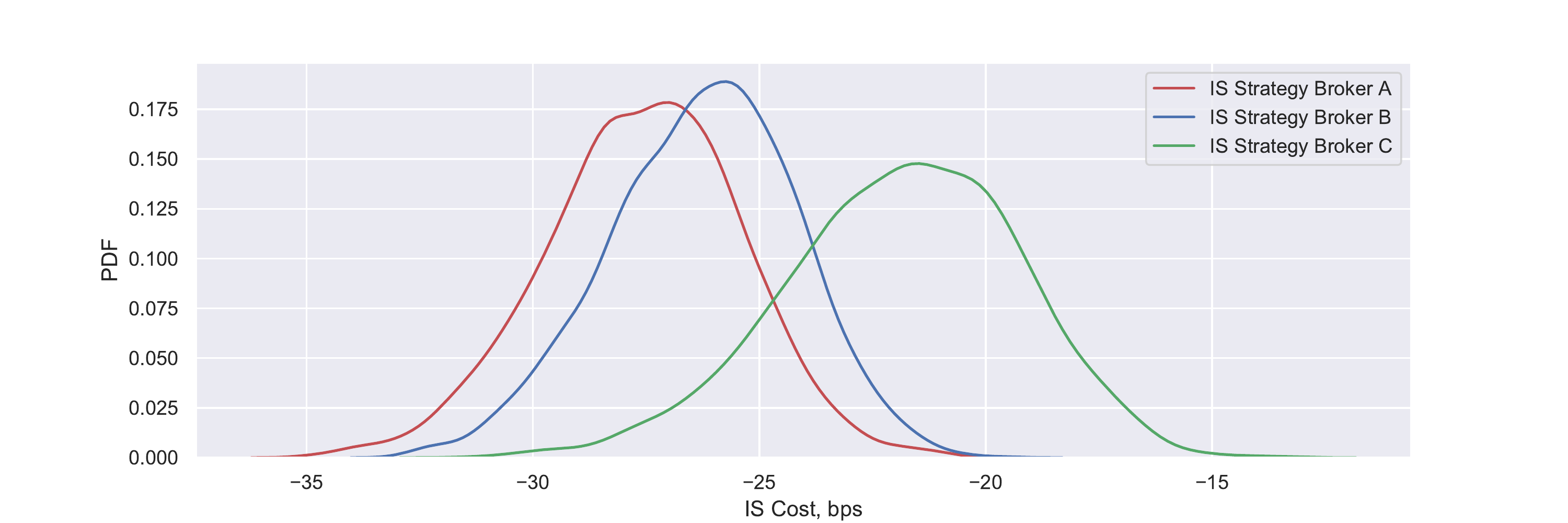}}\hspace{1em}%
  \subcaptionbox{Figure 5: The posterior distributions of the VWAP cost.}{\includegraphics[width=6in,height=2.5in]{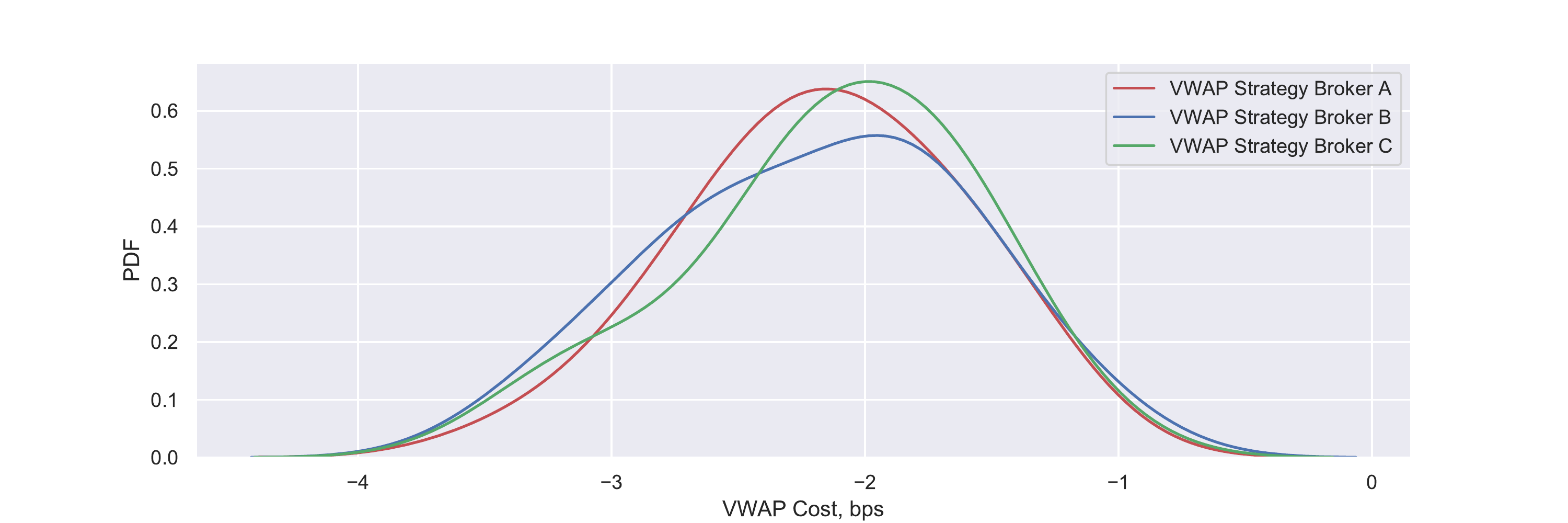}}\hspace{1em}%
  \hspace*{\fill}%
\end{figure}

\clearpage


\end{document}